\documentclass[prl,twocolumn,superscriptaddress,showpacs]{revtex4}
\def\beq{\begin{equation}}
\def\eeq{\end{equation}}
\def\beqn{\begin{eqnarray}}
\def\eeqn{\end{eqnarray}}
\usepackage{amsmath}
\usepackage{graphicx}
\usepackage{epsfig}
\usepackage[usenames]{color}

\begin{document}
\title{Confinement-induced Berry phase and helicity-dependent photocurrents}
\begin{abstract}
The photocurrent in an optically active metal is known to contain a component that switches sign with the helicity of the incident radiation.  At low frequencies, this current depends on the orbital Berry phase of the Bloch electrons via the ``anomalous velocity'' of Karplus and Luttinger.  We consider quantum wells in which the parent material, such as GaAs, is not optically active and the relevant Berry phase only arises as a result of quantum confinement.  Using an envelope approximation that is supported by numerical tight-binding results, it is shown that the Berry phase contribution is determined for realistic wells by a cubic Berry phase intrinsic to the bulk material, the well width, and the well direction.  These results for the magnitude of the Berry-phase effect suggest that it may already have been observed in quantum well experiments.
\end{abstract}

\author{J.~E.~Moore}
\affiliation{Department of Physics, University of California, Berkeley, Berkeley, CA 94720}
\affiliation{Materials Sciences Division, Lawrence Berkeley National Laboratory, Berkeley, CA 94720}
\author{J.~Orenstein}
\affiliation{Department of Physics, University of California, Berkeley, Berkeley, CA 94720}
\affiliation{Materials Sciences Division, Lawrence Berkeley National Laboratory, Berkeley, CA 94720}
\pacs{78.56.-a, 73.63.Hs, 78.67.De}
\maketitle

The ``Fermi liquid'' theory of metals developed by Landau captures the microscopic details of band structure and interactions in a small number of parameters that quantify how the quasiparticles in the metal differ from free electrons.  While this theory has been successful in a wide variety of materials, it is now considered to be incomplete: it does not include physics resulting from the Berry phase of the Bloch wavefunctions.  Berry phases result when a quantum-mechanical wavefunction changes smoothly as a function of some parameter~\cite{berry}.  The orbital Berry phases in metals cause an ``anomalous velocity'' of Bloch electrons, derived by Karplus and Luttinger~\cite{karplusluttinger} and later interpreted as a Berry-phase effect~\cite{sundaramniu,jungwirth}.

The Berry phase that influences basic transport properties vanishes in materials that are symmetric with respect to both inversion and time-reversal.  In metals that break time-reversal (TR) symmetry (ferromagnets and antiferromagnets), the Berry phase leads to an ``intrinsic'' mechanism of the anomalous Hall effect (AHE).  The other broad class of materials with nonzero Berry phase are those in which TR is preserved but inversion symmetry is broken; in insulators, these Berry phases underlie the modern theory of polarization~\cite{ksv,resta}.  Berry phases from spin-orbit coupling lead to   ``topological insulators'' in two~\cite{kane&mele-2005,zhangscience1,molenkampshort} and three~\cite{fu&kane&mele-2007,moore&balents-2006,hsiehshort} dimensions.

By contrast, at present there is no experimental observation that has been associated with the non-vanishing Berry phase that is expected in metals that break inversion symmetry.  However, the helicity-dependent photocurrent in an optically active metal contains a Berry-phase contribution, as shown recently by Deyo {\it et al.}~\cite{golub}, who give a semiclassical transport analysis of photocurrents at linear order in applied intensity.  We believe that this effect, also known as the circular photo-galvanic effect (CPGE), is quite fundamental and differs in important ways from previous Berry-phase phenomena: it is nonlinear, frequency-dependent, and controlled by the Berry-phase contribution at low frequency.


In this paper we compute the Berry-phase contribution to helicity-dependent photocurrents in realistic circumstances, in order to allow quantitative comparison with recent experiments on semiconductor quantum wells. In these systems, the Berry phase contribution vanishes in the bulk; the the photocurrent is generated by quantum confinement. We first give a simple derivation of the Berry-phase contribution in two dimensions in the relaxation-time approximation.  A microscopic tight-binding calculation for a model GaAs (110) quantum well is then described.  It yields a magnitude of the photocurrent that agrees well with experiments by the Regensburg group~\cite{ganichevprl,ganichev110,ganichevnew}. Finally, we show via an envelope-function approach that the confinement-induced Berry phase in a general zincblende quantum well can be parameterized in terms of only two numbers, one intrinsic to the bulk material and the other a geometrical factor determined by the surface orientation and well width.



Our starting point is the semiclassical equation for a wavepacket of Bloch electrons.  The velocity vector has two terms~\cite{karplusluttinger,sundaramniu,haldaneberry}.  One is the familiar semiclassical velocity, which is determined near a parabolic minimum by the effective mass, and the other results from the change in the spatial location of the electron within the unit cell as its wavevector ${\bf k}$ moves through the Brillouin zone:
\beq
{d x^a \over dt} = {1 \over \hbar} {\partial \epsilon_n({\bf k}) \over \partial {k_a}} + {\cal F}^{ab}_n({\bf k}){d k_b \over dt}.
\label{anomvel}
\eeq
In the second term, known as the anomalous velocity, the Berry flux is written with the same symbol ${\cal F}^{ab}$ as the electromagnetic field tensor in order to stress certain similarities.  (The band index $n$ is suppressed from now on as we assume a single spin-degenerate partially occupied band).  ${\cal F}^{ab}$ is the curl of a ``vector potential'' obtained by momentum-space derivatives of the periodic part of the electronic eigenstates $u_k({\bf x})$ (here $\partial_a = \partial_{k_a}$):
\beqn
{\cal F}^{ab}({\bf k}) &=& \partial^a {\cal A}^b({\bf k}) - \partial^b {\cal A}^a({\bf k}),\\
{\cal A}^a({\bf k}) &=& - i \langle u_k({\bf x}) | \partial_a u_k({\bf x}) \rangle.
\eeqn
As in electromagnetism, ${\cal A}^a$ is gauge-dependent (i.e., changes under a phase change of Bloch states) but ${\cal F}^{ab}$ is gauge-independent.  It is also antisymmetric, and time-reversal symmetry requires ${\cal F}^{ab}({\bf k}) = -{\cal F}^{ab}(-{\bf k}).$

In two dimensions, the Berry flux has only one nonzero component ${\cal F}^{12} = -{\cal F}^{21}$, which can be pictured as a vector ${\bf \Omega}({\bf k}) = {\cal F}^{12}({\bf k}) {\bf \hat z}$ pointing out of the 2D plane.  The anomalous velocity leads to a net current if ${\bf \dot k} \times {\bf \Omega}({\bf k})$ has a nonzero average over the electron distribution.  In the relaxation-time approximation, ${\bf \dot k}$ is constant over the electron distribution.  In the parabolic approximation, the ground-state distribution of electrons is circular.  Time-reversal symmetry implies that ${\bf \Omega}({\bf k})$ averages to zero in the ground-state electron distribution.

Even without a calculation, we can obtain some intuition for how the anomalous velocity will lead to helicity-dependent photocurrents if ${\bf \Omega}({\bf k})$ is proportional to $k_x$, which is the symmetry of the GaAs quantum well discussed below.
Consider a low-frequency circularly polarized incident wave whose electrical field ${\bf E}$ lies in the plane of the 2D system.  If the relaxation time $\tau$ is short compared to the period of the incident wave, so that $\omega \tau \ll 1$, then the instantaneous distribution in $k$-space is slightly displaced from the parabolic minimum, and this displacement circles with frequency $\omega$ around the origin (Fig.~1).  At an instant when $e {\bf E}$ points along $+{\bf \hat x}$, the average of ${\Omega}({\bf k})$ is along $+{\bf \hat z}$, while the electron distribution has ${\bf \dot k}$ along $\pm {\bf \hat y}$ according to the sense of circular polarization.  This gives a current directed along the $x$-axis, and the {\it same} direction of current is obtained after a half-period.

\begin{figure}
\includegraphics[width=3.0in]{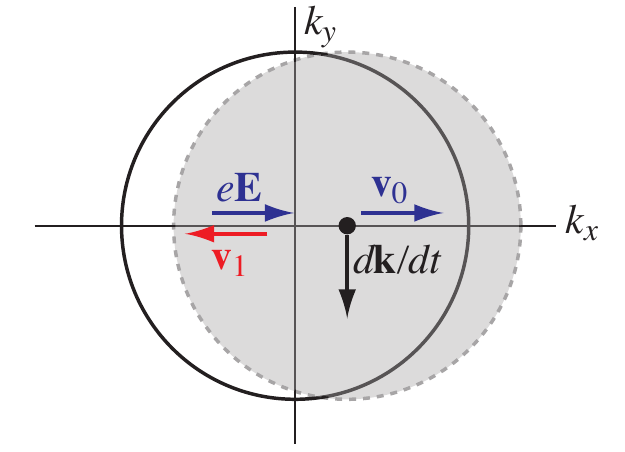}
\caption{Snapshot of electron distribution in a circularly polarized wave.  The electron distribution is pushed away from the ground state (solid circle) slightly for small $\omega \tau$.  At the time shown, the conventional velocity ${\bf v}_0$ and anomalous velocity ${\bf v}_1$ are in opposite directions.  After a half-period, the conventional velocity changes sign but the anomalous velocity does not.  The directions assume ${\bf \Omega} = \beta k_x {\bf \hat z}$ with $\beta > 0$.}
\end{figure}

More formally, the current density $\bf j$ arising from the anomalous velocity in (Eq.~\ref{anomvel}) is
\beq
\textbf{j}= e \int {d^2k\over 4\pi^2} [{\bf \dot k} \times {\bf \Omega}({\bf k})]g({\bf k}),
\eeq
where $g({\bf k})$ is the deviation from the ground state conduction electron distribution. Within the relaxation time approximation with isotropic mean-free time $\tau$, we have
\beq
g({\bf k}) = \left({{e {\bf E} \cdot  {\bf v}_n({\bf k})} \over {1/\tau - i \omega}}+c.c. \right)\left({\partial f \over \partial \epsilon}\right)_0
\eeq
and ${\bf \dot k} = {e {\bf E}/\hbar} +c.c.$ (As in the AHE~\cite{jungwirth}, impurity fields are not included in the anomalous velocity.)  Here ${\bf v}_n({\bf k})$ is the normal velocity $\nabla_k \epsilon({\bf k})$, and the partial derivative of the Fermi function $f$ with respect to energy $\epsilon$ is evaluated at the chemical potential. The optical field ${\bf E}$ is specified in phasor notation: ${\bf E}(t)= (E_x {\bf \hat x} + E_y {\bf \hat y})e^{i\omega t}$.  The Berry flux is taken to be ${\Omega}({\bf k}) = \beta k_x {\bf \hat z}$ for some constant $\beta$ with units of volume, because a Berry flux linear in ${\bf k}$ is required for the effect to appear at quadratic order, and any linear combination of $k_x$ and $k_y$ can be brought to the above by a rotation.  For a circular Fermi surface and low enough frequency that the semiclassical equation is valid, the \emph{dc} current density $\textbf{j}_{dc}$ is the real part of
\beqn
{\bf{\tilde{j}}} &=& {\beta e^3 \over 4 m \pi^2} {i \omega ({\bf E} \times {\bf \hat z}) \over 1/ \tau^2 + \omega^2}  \int d^2k\,k_x\,({\bf E}^* \cdot {\bf k}) \left({\partial f \over \partial \epsilon}\right)_0.
\label{curreq}
\eeqn
Switching to polar coordinates, the ${\bf k}$ integral becomes
\beqn
&&\int k^3\,dk\,d\theta\, (\cos \theta) (E_x^* \cos \theta + E_y^* \sin \theta) \left({\partial f \over \partial \epsilon}\right)_0
\cr &=& {2 \pi m^2 \over \hbar^4}  E_x^* \int \epsilon\,d\epsilon\,\left({\partial f \over \partial \epsilon}\right)_0 = -{2 \pi m^2 \over \hbar^4}  E_x^* \epsilon_F,
\eeqn
where $\epsilon_F$ is the Fermi energy.  Inserting this result into the complex current density (Eq.~\ref{curreq}) gives
\beqn
\textbf{j}_{dc} &=& {\beta n e^3 \over 2\hbar^2} {1 \over 1/\tau^2 + \omega^2}\Big[ i\omega (E_x E_y^* - E_y E_x^*){\bf\hat x} \cr
&&+1/\tau(E_x E_y^* + E_y E_x^*){\bf\hat x}+|E_x|^2 {\bf\hat y}\Big].
\label{berryeq}
\eeqn
In Eq.~\ref{curreq} we substituted $\epsilon_F = \pi\hbar^2 n /m$ where $n$ is the electron areal density and restored a factor of 2 for spin.

The first two terms on the right-hand side of Eq.~\ref{berryeq} are Berry-phase contributions to the circular (CPGE) and the linear (LPGE) photogalvanic effects, respectively.  The LPGE, in which the photocurrent is maximal for linearly polarized light, is allowed in all acentric media. The CPGE, which is maximal for circular polarization and changes sign with helicity of the light wave, is allowed in the subset of acentric point groups that are optically active, or "gyrotropic."  The last term on the right-hand side of (Eq.~\ref{berryeq}) is a more typical form of photovoltaic effect, in which the asymmetry of the medium fixes the sign of the photocurrent, independent of the polarization state of the light.  But all these effects differ fundamentally from the photovoltaic effect in a solar cell, for example, where the photocurrent is just determined by the number of absorbed photons; here all terms depend on the polarization state of the incident light.

Doped semiconductors are attractive systems in which to search for the above effects because of their relatively long relaxation times.  However, zincblende semiconductors have too much symmetry to allow the above effects in bulk. Instead, $\beta$ must be generated by a confinement potential such as arises in a quantum well structure.  We will compute this type of Berry phase by two approaches and compare the results to photocurrent experiments.

The first approach is an explicit microscopic calculation on model quantum wells.  Consider the lowest conduction subband of a (110) quantum well in GaAs and assume that the crystal structure (Fig.~2a) is unmodified by creation of the well, which then is simply an additional electrical potential.  The spacing between (110) planes is $\sqrt{2} a / 4$ = 2.00 \AA\ with lattice constant $a$ = 5.65 \AA.
The maximal remaining point group symmetry is $C_{2v}$: there are two mirrors, in (110) and $(1{\bar 1}0)$ planes, and a $\pi$ rotation around [001] that is the product of the mirrors.  An asymmetric well potential eliminates the (110) mirror and the rotation, reducing the symmetry to $C_s$.  Either symmetry allows a nonzero Berry phase parameter $\beta$.

Our computational approach starts from reference tight-binding parameters for conduction and valence bands in GaAs~\cite{harrisonbook}.  It is known that this tight-binding model gives a roughly correct band structure, but overestimates the band gap and effective mass. Achieving agreement with experimental bandgaps and masses requires correlation methods~\cite{hybertsenlouie} that are difficult to implement on a quantum well of realistic size, although the envelope expression below explains how these could be applied.  Tight-binding results for the Berry phase may be more accurate than those for the band gap, as the former results purely from geometrical properties of the wavefunctions.  Spin-orbit coupling is ignored since in the conduction band it is weak compared to orbital effects.

\begin{figure}
\includegraphics[width=3.25in]{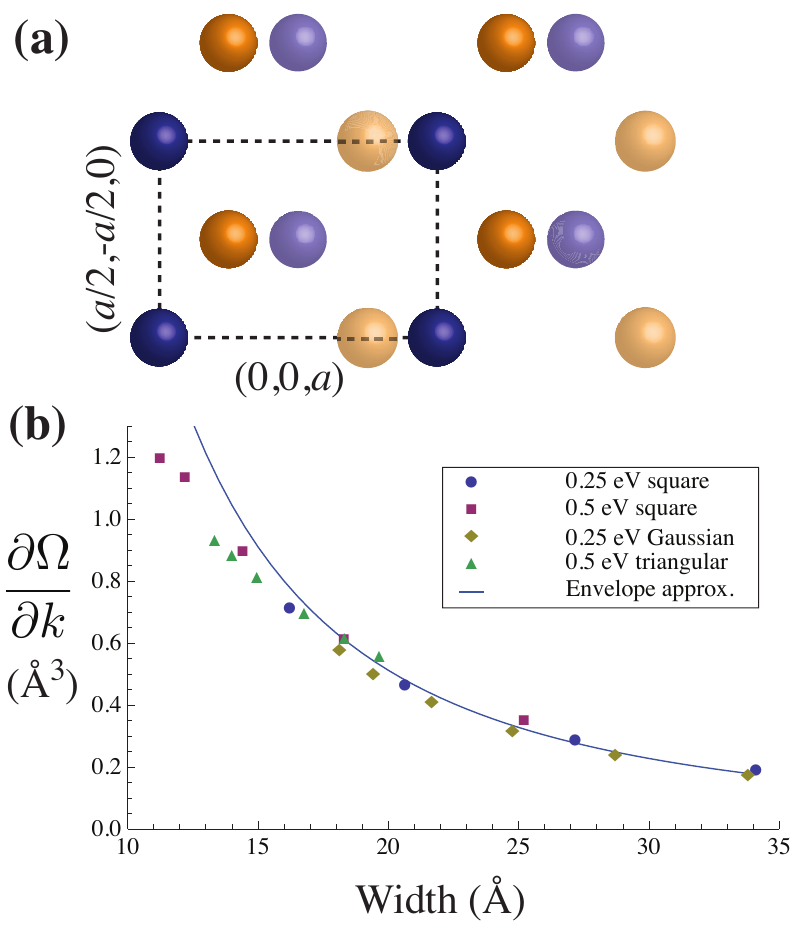}
\caption{Berry phase in (110) quantum wells.  (a) Positions of atoms (Ga gold, As blue) within (110) planes of GaAs showing inversion asymmetry.  Shadowed atoms are in the next plane above or below.  
(b) Strength of Berry phase factor $\beta = {\partial \Omega / \partial k}$ for (110) quantum wells: square wells of depth 0.25 eV and 0.5 eV, Gaussian well of depth 0.25 eV, and triangular well of depth 0.5 eV.  The wavefunction width $2 \sqrt{\langle {\tilde x}^2 \rangle}$ (horizontal axis), ${\tilde x} = (x+y)/\sqrt{2}$ changes with well potential and determines $\beta$ almost independently of well shape.
}
\end{figure}

The resulting Berry phase factors for a variety of quantum well potentials are given in Fig.~2b.  We find that $\beta$ is on the order of 1 ${\AA}^3$ and decreases with increasing well width.  For comparison to experiment, consider the current $J = j_x L$ from optical power $P$ illuminating a square of side $L$.  The CPGE current for circular polarization is
\beq
J = G \left( {4 \pi \alpha \beta m P \over e \hbar L} \right) {\omega \tau \over 1+ \omega^2 \tau^2},
\eeq
where $n$ was eliminated in favor of the Drude conductance $G = n e^2 \tau / m$ and $\alpha = e^2 / (4 \pi \epsilon_0 \hbar c) \approx 1/137$ is the fine structure constant.  The quantity in parentheses can be viewed as an effective voltage that combines with the Drude conductance. From the above formula, $P$=1 W incident on a 1 mm$\times$1 mm area of quantum well with Drude conductance 10$^{-2}$ mho leads to a peak current of 1.16 nA at $\omega \tau = 1$, where we have taken $\beta=1$ ${\AA}^3$ and used the effective mass $0.067\ m_e$ of conduction electrons in GaAs. Currents of this magnitude are found by the Regensburg group in experiments on doped semiconductor quantum wells and heteroepitaxial interfaces~\cite{ganichevprl,ganichev110,ganichevnew}.

Extending the tight-binding results given above to a broader range of experiments requires understanding the confinement-induced Berry phase in an arbitrary quantum well.  We choose an approach similar to the envelope approximation for energy bands in a quantum well~\cite{bastardbookshort}.  A three-dimensional zincblende semiconductor like GaAs is acentric but nevertheless has no Berry-phase photocurrent at first order in the applied intensity.  Microscopically, near the Brillouin zone origin $\Gamma$ the remaining symmetries restrict the form of the Berry flux to be in the $T_1$ representation; ${\bf \Omega}$ is not linear in $k$ near $\Gamma$ but instead
\beq
{\bf \Omega} = \lambda \left(k_x (k_y^2-k_z^2), k_y (k_z^2-k_x^2),k_z (k_x^2-k_y^2)\right).
\eeq  The constant $\lambda$ determines Berry-phase effects in the material when the Fermi surface is near $\Gamma$.

A quantum well can reduce the symmetry and combine with $\lambda$ to allow a linear photocurrent.
Results on the (110) well (Fig.~2) suggest that the detailed form of the confined state is not too important, so we take a Gaussian wavepacket of conduction band states:
\beq
\psi_{2D}(x,y,z)
= C \int e^{-k_z \sigma^2} e^{i {\bf k} \cdot {\bf r}} u_{\bf k}(x,y,z)\,dk_z.
\eeq
This wavefunction combines the intra-unit-cell dependence $u$ of the conduction band with a Gaussian envelope on longer scales.  $C$ is a normalization constant.  (For simplicity, we take coordinates with ${\bf \hat z}$ perpendicular to the well plane and later account for the rotation relative to the crystalline unit cell.) The wavefunction is localized in the ${\bf \hat z}$ direction, with $\langle z^2 \rangle = \sigma^2$ for $\sigma$ much larger than a unit cell, which is the regime where the envelope approximation is applicable.

The above wavefunction is a 2D Bloch state, and the 2D Berry flux $F^{\rm 2D}_{xy}$ arises from the 3D Berry flux of the original crystal.  The envelope result is easily stated: the 2D Berry flux averages the 3D Berry flux over $k_z$,
\beq
F^{\rm 2D}_{xy}(k_x,k_y) = {\int e^{-2 \sigma^2 k_z^2} F^{\rm 3D}_{xy}(k_x,k_y,k_z) \,dk_z \over
\int e^{-2 \sigma^2 k_z^2} \,dk_z}.
\eeq
For the (110) well discussed above, inserting the 3D Berry phase strength $\lambda$ to compute the 2D Berry-phase coefficient $\beta$ gives a simple function of $\lambda$, the well width $\sigma$, and a dimensionless factor $g_0$ set by well direction:
\beq
\beta = {g_0 \lambda \over \sigma^2} = - {\lambda \over 8 \sigma^2}, \label{berryfactor}
\eeq
which is the curve shown in Fig.~2 with $\lambda = -410 \AA^5$ calculated from our {\it bulk} tight-binding wavefunctions.

The same approach to compute the confinement-induced Berry phase can be applied to other materials and well geometries.  For example, the geometrical factor of $g_0 = -1/8$ in (Eq.\ref{berryfactor}) is modified in a $(11n)$ well to
\beq
g_n = {n^2 - 1 \over (2 n^2 + 4)^{3/2}}.
\eeq
The vanishing at $n=1$ is expected because a $(111)$ quantum well has an extra symmetry (a 3-fold rotation axis).

If the Berry-phase mechanism of photocurrents in quantum wells can be confirmed experimentally, Berry-phase phenomena will have been observed in all the four basic materials classes, magnetic and nonmagnetic metals and insulators.  However, just as in the case of the AHE in TR-breaking metals, other symmetry-allowed mechanisms coexist with the Berry-phase contribution. Only after many years of debate has it been established that the Berry-phase mechanism for AHE is dominant in certain regimes of temperature and disorder.

We believe the Berry phase contribution to photocurrent in acentric metals can be identified more easily, through its signature dependence on the frequency and polarization state of the light.  For example, Eq. \ref{berryeq} predicts that the CPGE and LPGE effects are linked; the Berry-phase generates both with equal magnitude and they vary with frequency like the imaginary and real parts of the Drude conductance, respectively~\cite{golub}. Moreover, the Berry phase mechanism is readily distinguished from alternate mechanisms involving inter-subband effects (either spin or orbital) by its dependence on the angle of incidence, $\theta$, of the light.  While the Berry-phase related currents are driven only by in-plane components of the optical field, the inter-subband mechanisms are driven by $E_z$ as well. Thus the Berry-phase photocurrents vary strictly as $\cos \theta$, while inter-subband response has terms proportional to both $\cos \theta$ and $\sin \theta$.  In this regard, the early measurements of CPGE in a (113) quantum well by Ganichev {\it et al.}~\cite{ganichevprl} are highly suggestive; although both $\cos \theta$ and $\sin\theta$ dependences are allowed by symmetry, only the $\cos \theta$ variation is observed.

Finally, we suggest that the envelope approximation developed here can be applied to confinement-induced Berry-phase phenomena in a wide variey of heterojunctions and quantum wells.
Beyond interest in the effect \emph{per se}, measurements of helicity-dependent photocurrents could be used as a means to obtain the leading non-zero Berry phase in many materials, in the same way as other transport measurements are used to obtain the parameters of conventional Fermi liquid theory.

The authors acknowledge helpful conversations with J. Folk and A. MacDonald and support from NSF DMR-0804413 (JEM) and DOE BES (JO).

\end{document}